\documentclass[%
 aip,
 amsmath,amssymb,
 reprint,%
]{revtex4-1}

\usepackage{graphicx}
\usepackage{dcolumn}
\usepackage{bm}
\usepackage{color}

\usepackage[utf8]{inputenc}
\usepackage[T1]{fontenc}
\usepackage{mathptmx}
\usepackage{etoolbox}
\usepackage{multirow}

\makeatletter
\def\@email#1#2{%
 \endgroup
 \patchcmd{\titleblock@produce}
  {\frontmatter@RRAPformat}
  {\frontmatter@RRAPformat{\produce@RRAP{*#1\href{mailto:#2}{#2}}}\frontmatter@RRAPformat}
  {}{}
}%
\makeatother
\begin{document}

\title{In-situ tunable interaction with an invertible sign between a fluxonium and a post cavity}

\author{D. G. Atanasova}
\affiliation{University of Innsbruck, Institute for Experimental Physics, 6020 Innsbruck, Austria}
\affiliation{Institute for Quantum Optics and Quantum Information, Austrian Academy of Sciences, 6020 Innsbruck, Austria}

\author{I. Yang}
\affiliation{University of Innsbruck, Institute for Experimental Physics, 6020 Innsbruck, Austria}
\affiliation{Institute for Quantum Optics and Quantum Information, Austrian Academy of Sciences, 6020 Innsbruck, Austria}
\author{T. Hönigl-Decrinis}
\affiliation{University of Innsbruck, Institute for Experimental Physics, 6020 Innsbruck, Austria}
\affiliation{Institute for Quantum Optics and Quantum Information, Austrian Academy of Sciences, 6020 Innsbruck, Austria}

\author{D. Gusenkova}
\affiliation{IQMT, Karlsruhe Institute of Technology, 76131 Karslruhe, Germany
}
\affiliation{PHI, Karlsruhe Institute of Technology, 76131 Karlsruhe, Germany}

\author{I. Pop}
\affiliation{IQMT, Karlsruhe Institute of Technology, 76131 Karslruhe, Germany
}
\affiliation{PHI, Karlsruhe Institute of Technology, 76131 Karlsruhe, Germany}
\affiliation{Physics Institute 1, Stuttgart University, 70569 Stuttgart, Germany}

\author{G. Kirchmair}
\affiliation{University of Innsbruck, Institute for Experimental Physics, 6020 Innsbruck, Austria}
\affiliation{Institute for Quantum Optics and Quantum Information, Austrian Academy of Sciences, 6020 Innsbruck, Austria}

\date{\today}

\begin{abstract}
Quantum computation with bosonic modes presents a powerful paradigm for harnessing the principles of quantum mechanics to perform complex information processing tasks. In constructing a bosonic qubit with superconducting circuits, nonlinearity is typically introduced to a cavity mode through an ancillary two-level qubit. However, the ancilla's spurious heating has impeded progress towards fully fault-tolerant bosonic qubits. The ability to in situ decouple the ancilla when not in use would be beneficial but has so far only been realized with tunable couplers or additional parametric drives. This work presents a novel architecture for quantum information processing, comprising a 3D post cavity coupled to a fluxonium ancilla via a readout resonator. This system's intricate energy level structure results in a complex landscape of interactions whose sign can be tuned in situ by the magnetic field threading the fluxonium loop without the need of additional elements. Our results could significantly advance the lifetime and controllability of bosonic qubits.
\end{abstract}

\maketitle
    
Among the various platforms explored for quantum computing, bosonic qubits represent a compelling approach. Unlike traditional qubit systems, which use two-level systems, bosonic qubits leverage the infinite-dimensional Hilbert space of harmonic oscillators, allowing for the encoding of quantum information in more complex and potentially more robust ways. Encoding techniques such as GKP (Gottesman-Kitaev-Preskill)\cite{GKP_2001,Phillip_2020,Sivak_break_even_2023}, cat states\cite{ofek_extending_2016,grimm_stabilization_2020}, kitten\cite{ni_break_even_2023}
and binomial states\cite{Binomial_2016} have already demonstrated how bosonic qubits can store and process information with high fidelity in a single harmonic oscillator. Similarly, there are schemes for distributing information across two bosonic modes i.e. dual-rail erasure qubits\cite{teoh_dual-rail_2023}. All these methods can significantly reduce the detrimental effect of the inherent loss mechanisms in the system, resulting in qubits with enhanced storage capabilities and reduced hardware control stack.   

Superconducting microwave cavities are a great candidate to host bosonic qubits as their lifetimes are enhanced by at least an order of magnitude compared to conventional junction-based superconducting qubits, such as transmons and fluxonia. To control such a bosonic mode, a nonlinearity is typically introduced to the cavity through a transmon qubit\cite{Koch_transmon_2007}, referred to as ancilla. However, it has been shown that this ancilla is the main limitation of the bosonic qubit due to residual excitations out of the computational space and its chaotic behavior under strong drives\cite{Cohen_transmon_chaos_2023}. This motivated a search for alternatives: removing the transmon overall\cite{reglade_cat_2024} or using an ancilla bosonic qubit \cite{Royer_kerr_ancilla_2020}. While both approaches succeed in their objective, they come with disadvantages. The former uses an increased control complexity, i.e. a bigger instrument control stack is needed, while the latter suffers from weaker interaction strengths leading to longer experiment times. A perfect replacement would be an ancilla that can be decoupled selectively from the bosonic mode after the state preparation and tomography. So far, this has been realized by introducing either additional parametric drives\cite{Rosenblum_2018,Koottandavida_2024} or  elements such as tunable couplers\cite{noh2021strongparametricdispersiveshifts}.

In this work, we couple a $\lambda/4$ high Q post cavity to a fluxonium ancilla qubit\cite{Manucharyan_fluxonium_2009} via its readout resonator. The complex energy landscape reveals novel transversal interactions between the bosonic mode and its ancilla, which can be tuned in situ between negative and positive values smoothly through zero. This system allows for both, sufficient interaction strength required for the control of the storage mode and for decoupling the high-Q cavity from the fluxonium, removing the main source of dephasing of the bosonic qubit. 

\section{Experimental Setup}
\begin{figure}[t]
 \centering
\includegraphics{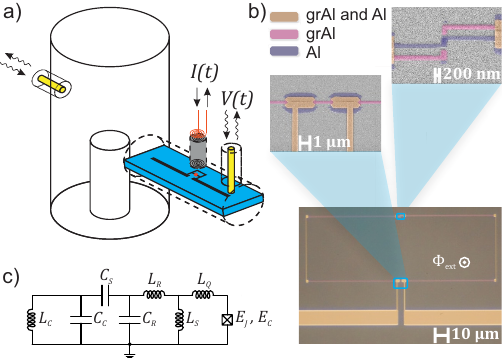}
\caption{\label{fig:setup} Experimental setup. (a) The device consists of a high Q cavity, fluxonium (the junction is marked with a red cross), and readout resonator on a sapphire chip (blue), flux hose (grey), and superconducting coil (red), and coupling pins (yellow). (b) False-colored scanning electron microscope (SEM) image of the sample around the fluxonium. The colors correspond to the thin-film material; violet for aluminum (\text{Al}), pink for granular aluminum (\text{grAl}), and orange for overlapped \text{grAl} and \text{Al} material.  (c) Lumped-element model of the setup. The cavity ($L_C, C_C$) is capacitively coupled ($C_S$) to the resonator ($L_R, C_R$) which in turn is inductively coupled ($L_S$) to the fluxonium qubit consisting of a Josephson junction ($E_J, E_C$) shunted by a superinductance ($L_Q$).}
\end{figure}

In our experiment, schematically shown in Fig. \ref{fig:setup}a, an aluminum post-cavity serves as the memory mode. The control is provided by a fluxonium\cite{grunhaupt_granular_2019,gusenkova_quantum_2021} (Fig. \ref{fig:setup}b) consisting of an aluminum-aluminum oxide-aluminum (\text{Al}-\text{AlOx}-\text{Al}) Josephson junction (Fig. \ref{fig:setup}b right inset), shunted by a granular aluminum (\text{grAl}) superinductance (fabrication details in S2). The fluxonium is inductively coupled to a readout resonator via a shared \text{grAl} section (Fig. \ref{fig:setup}b, left inset). The readout resonator shares a capacitance with the cavity. The additional control knob for the magnetic bias of the fluxonium is provided by a small superconducting coil sitting on top of a home-built magnetic flux hose \cite{gargiulo_flux_hose_2021}, which guides the field into the superconducting enclosure. 

The fundamental mode of the coaxial cavity sits at frequency $\omega_{\text{C}}/2\pi \approx 4.5$ GHz, much lower than the cut-off frequency of the waveguide section above it, resulting in its high-quality factor \cite{reagor_reaching_2013,heidler_highQ_cavity_2021}. The measured lifetime of this mode with the qubit coupled to it is $T_1^C=210\pm40\,\mu\text{s}$ (see appendix \ref{app:methods}). According to finite element simulations, the higher cavity modes start at about $13.37$ GHz. These modes spread more into the waveguide section giving them lower quality factors as they have a higher participation ratio with the lossy seam at the top of the cavity. The qubit is tunable in the range $\omega_Q/2\pi\in[1.02,13.45]$ GHz, meaning it can also couple to the higher cavity modes (see Appendix \ref{app:sim}). 

The short section of \text{grAl}, a high kinetic inductance material\cite{grünhaupt_2018_gral_loss}, serves as a mutual inductance between the qubit and the $\lambda/2$ readout resonator. The first resonator mode has a current antinode at the center, leading to higher participation of the kinetic inductance, while the second and all other even resonator modes have a current node at the same spot and the participation of the granular aluminum piece is negligible. Consequently, the qubit couples only to the fundamental readout resonator mode at $\omega_R/2\pi\approx6.8$ GHz and all other odd modes if we assume pure inductive coupling and no spurious capacitive coupling. Both the fundamental cavity and resonator modes are coupled to the control instruments via SMA pins with rates $\kappa^\text{ext}_C/2\pi=233\pm7$ Hz and $\kappa^\text{ext}_R/2\pi=549\pm 2$ kHz respectively.

\section{System Model}
The system Hamiltonian is comprised of three subsystems, namely the fluxonium $\hat{H}_{\text{Q}}$, the readout resonator $\hat{H}_{\text{R}} $ and the high Q cavity $\hat{H}_{\text{C}}$ connected by their respective interactions
$\hat{H} = \hat{H}_{\text{Q}} + \hat{H}_{\text{R}} + \hat{H}_{\text{C}} + \hat{H}_{\text{QR}} +\hat{H}_{\text{RC}}$. The fluxonium has the Hamiltonian
\begin{eqnarray}
\hat{H}_{\text{Q}} = 4E_C\hat{n}^2 + \frac{E_L}{2}\left(\hat{\varphi}+\varphi_{\text{ext}}\right)^2 -E_J\cos{(\hat{\varphi})},
\label{eq:fluxonium_qubit_hamiltonian}
\end{eqnarray} 
where $\hat{n}$ is the charge number operator, $\hat{\varphi}$ is its conjugate  phase operator and $\varphi_\text{ext} = 2\pi\Phi_\text{ext}/\Phi_0$ is the reduced external flux with $\Phi_0=h/2e$ being the flux quantum. The capacitive energy can be written in terms of the total qubit capacitance, which is dominated by the junction, $E_C = e^2/2C_Q$, while the inductive energy contains the contribution of the \text{grAl} superinductance and the shared \text{grAl} inductance $E_L=\phi_0^2/(L_Q+L_S)$, where $\phi_0=\Phi_0/2\pi$ is the reduced flux quantum. $E_J$ is the Josephson energy of the junction $E_J=\phi_0I_c$, with $I_c$ being the critical current of the junction. 

Determining the full system Hamiltonian requires different approaches depending on the physical dimensions of its constituents compared to the wavelengths of interest. If the whole system would satisfy the lumped element approximation\cite{pozar_microwave_2012}, an equivalent circuit model can be drawn (as shown in Fig. \ref{fig:setup}c) and a full Hamiltonian diagonalization is possible\cite{clark_fluxonium_2016}. However, this method is unsuitable in our case as the post cavity is a distributed element. An ideal way to quantize distributed element circuits is black box quantization (BBQ)\cite{nigg_BBQ_2012} but it necessitates only weakly anharmonic modes, incompatible with the fluxonium. Instead, we solve the Hamiltonian in the normal mode basis for the resonator and the cavity, i.e. $\hat{H}_R=\hbar\omega_R \hat{a}^\dagger\hat{a}$ and $\hat{H}_C=\hbar\omega_C \hat{b}^\dagger\hat{b}$ with $\omega_R, \omega_C$ being the fundamental mode frequencies of the resonator and the cavity, respectively, and $\hat{a}, \hat{b}$ - their annihilation operators respectively. The inductive coupling between the qubit and the resonator is modeled as  $\hat{H}_\text{QR} = -\hbar g_\text{QR}\hat{\varphi}(\hat{a}^\dagger+\hat{a})$, where $g_\text{QR}$ is the interaction strength. Additionally, the capacitive coupling between the resonator and the post cavity is  $\hat{H}_\text{RC} = \hbar g_\text{RC}(\hat{a}^\dagger-\hat{a})(\hat{b}^\dagger-\hat{b})$ with the interaction strength $g_\text{RC}$.

\begin{figure*}
\includegraphics{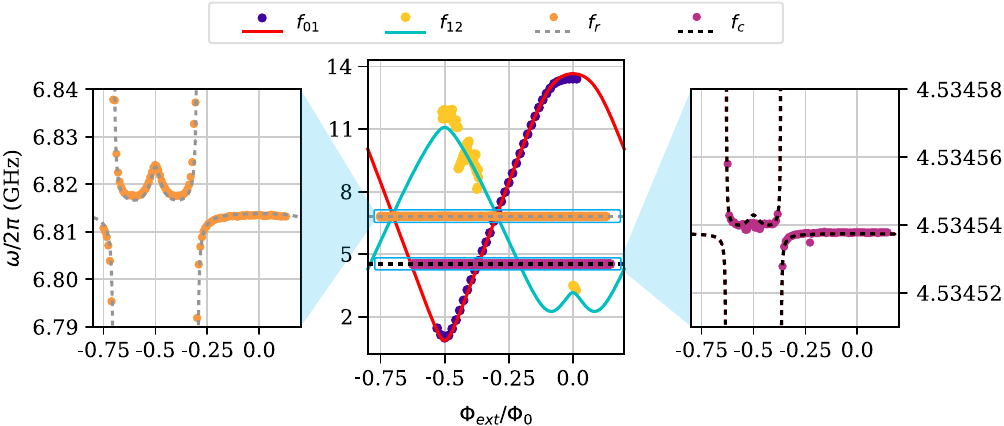}
\caption{\label{fig:specs} System spectrum. (center) The qubit frequency is measured as a function of flux bias. By fitting this spectrum we obtain the qubit energies $E_C/h=3.5$ GHz, $E_L/h=1.014$ GHz, and $E_J/h= 10.8$ GHz. The measured data for the fundamental transition is plotted in purple, together with its fit in red, the higher transition $f_{12}$ is in yellow with the simulation for it in cyan. (left) and (right) The readout resonator and high Q cavity spectra are measured as a function of flux respectively. The qubit is tuned through both the readout resonator and high Q cavity resulting in avoided crossings with coupling strengths $g_\text{QR}/2\pi\approx25.2$ MHz and $g_\text{QC}/2\pi\approx1.5$ MHz. The dashed lines correspond to their fits.
}
\end{figure*}
        
We calculate the energy eigenspectrum and the dispersive shifts of the system Hamiltonian with a numerical toolbox (scQubits \cite{Groszkowski_scqubitspython_2021}). The system model is fitted to experimental data obtained through qubit spectroscopy (Fig. \ref{fig:specs}). From the fit, we first acquire the relevant fluxonium energies, $E_C$, $E_L$, and $E_J$ in Eq. \ref{eq:fluxonium_qubit_hamiltonian}. Subsequently, we construct the Hilbert space by adding the relevant harmonic oscillators and their respective interactions. This step does not alter the fluxonium spectrum significantly except in the regions where the levels of the qubit and the harmonic oscillators cross. This fitting workflow gives us the full interaction Hamiltonian. With this information, we can predict the dispersive shifts between the subsystems as a function of the flux threading the fluxonium qubit. The complete analysis and details of the simulation methods can be found in appendix \ref{app:sim}.

\section{Experimental Results}

The qubit frequency is measured with two-tone spectroscopy. We monitor the resonator resonance 
while sweeping the frequency of a second tone with a signal generator. At the qubit frequency, the resonator shifts, and our readout signal changes. We repeat the procedure for different applied magnetic fields and fit each data to a Lorentzian curve to extract the qubit resonance frequency. The higher transition $f_{12}$ is extracted in the same way around $\Phi_\text{ext}/\Phi_0=-0.5$ where the non-zero excited state population of the qubit allows direct excitation to the next higher level. We fit this data with scQubits (Fig. \ref{fig:specs} center) and obtain the qubit energies $E_C/h=3.5$ GHz, $E_L/h=1.014$ GHz, and $E_J/h= 10.8$ GHz. The readout resonator and the cavity frequencies were extracted from direct spectroscopy measurements. The Hamiltonian is then constructed by including the harmonic oscillators and the interactions. The numerical results of the diagonalization are manually fitted to the cavity and resonator spectroscopy (Fig. \ref{fig:specs} right and left respectively) as a function of the external magnetic field. The avoided crossings between the resonator and the qubit, and between the cavity and the qubit give us the interaction strengths $g_\text{QR}/2\pi\approx25.2$ MHz with accuracy of about 10\% and  $g_\text{RC}/2\pi\approx8$ MHz with 15\%. The splitting of $2g_\text{QC}/2\pi\approx 3$ MHz we expect from the avoided crossing in Fig. \ref{fig:specs} (right) is a measure of the effective coupling rate between the qubit and the cavity, which is mediated by the resonator and determined by both $g_\text{QR}$ and $g_\text{RC}$. 

Near the high sweet spot $\Phi_\text{ext}/\Phi_0=0$, there is a discrepancy between the model and the measured data. A possible explanation is the cavity's higher modes which we expect around $13.37$ GHz, $13.73$ GHz, and $13.78$ GHz. Each of those modes will couple to the qubit via the readout resonator. In addition, the symmetry of the second additional mode will allow it to leak into the tunnel of the qubit and couple directly with the fluxonium. Owing to all the additional interactions at those frequencies, the model fails to fully capture the qubit behavior at the high frequencies. For further discussion on the limitations of this model, we refer the reader to appendix \ref{app:sim}.

\begin{figure}[h]
 \centering
\includegraphics{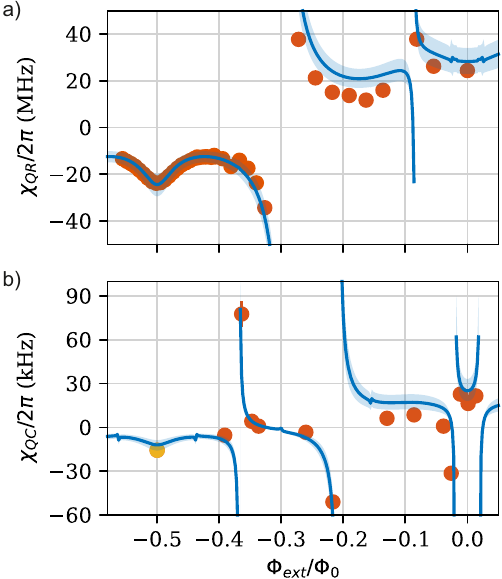}
\caption{\label{fig:disp_shifts} System dispersive shifts. (a) The dispersive shift between the qubit and readout resonator as a function of flux bias. The shaded region corresponds to a 10\% variation in the coupling strength $g_\text{QR}$. (b) The interaction between qubit and high Q cavity as a function of flux bias. The yellow point is measured with an echoed conditional displacement sequence, while the rest (red) are extracted from qubit spectroscopy as a function of photons in the cavity. The shaded region is a confidence interval, which corresponds to 15\% uncertainty of the coupling strength $g_\text{RC}$. Note the smooth transition between positive and negative values of the interaction around $\Phi_\text{ext}/\Phi_0 = -0.35$.}
\end{figure}
With the fits, the dispersive shift between the qubit and the resonator $\chi_{QR}$ and between the qubit and the cavity $\chi_{QC}$ can be calculated as depicted in Fig. \ref{fig:disp_shifts}. We extract the dispersive shift $\chi_{QR}$ from two-tone spectroscopy on the resonator, while the $\chi_{QC}$ measurements were performed in the time domain. The different methods are required as $\chi_{QR}$ is higher than the qubit linewidth at any external flux and $\chi_{QC}$ is lower.

Large displacements on the cavity enable the extraction of the Hamiltonian parameters\cite{eickbusch_ECD_2022} even when the dispersive shift is smaller than the qubit decoherence rate. Due to the cavity coupling to the qubit, the displaced coherent state will start rotating in phase space around the origin. After a fixed time, we displace the state back to the vacuum to obtain the acquired phase. The rotation frequency of the cavity state about the origin will be different for ground and excited qubit states and will depend on the dispersive shift (see Appendix \ref{app:methods} for more details). With this procedure, we obtain the dispersive shift at the low sweet spot (yellow data point in Fig. \ref{fig:disp_shifts}b). The other points are extracted from qubit spectroscopy as a function of photons in the cavity.

 The extracted data is plotted together with the numerical model predictions in Fig. \ref{fig:disp_shifts}. We notice that in Fig. \ref{fig:disp_shifts}b, the dispersive shift $\chi_{QC}$ is smoothly tunable between positive and negative values through zero, meaning we can switch off the interaction between the fluxonium and the cavity. The largest negative dispersive shift we measured is $\text{max}(\chi_{QC})/2\pi =-51 \pm 2$kHz, which is comparable to other recent works \cite{eickbusch_ECD_2022, Sivak_break_even_2023,ni_break_even_2023}. Also, the closest to zero we measured is $\text{min}(\chi_{QC})/2\pi=900\pm350$ Hz. While such measured results can also be achieved with a tunable transmon as an ancilla\cite{valadares_demand_2024}, in this setup it is possible to tune the interaction strictly to zero. Furthermore, we can tune to the opposite sign of $\chi_{QC}$, with our largest measured value at $78\pm9$ kHz, similar to tunable transmon implementations using additional elements like inductive coupler\cite{Chen_ind_coupler_2014}.
 
\begin{figure}[h!]
 \centering
\includegraphics{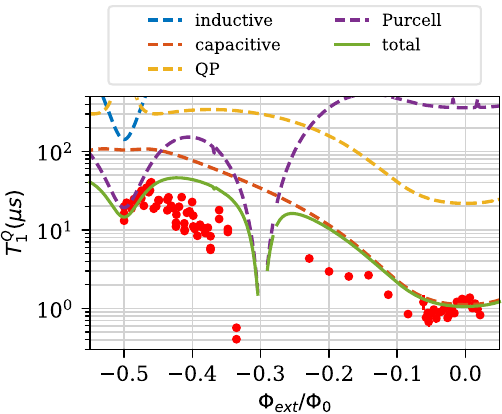}
\caption{\label{fig:t1} Fluxonium lifetime and loss mechanisms. The measured fluxonium $T_1^Q$ is plotted over the lifetime limits set by different loss channels: dielectric, inductive, non-equilibrium quasiparticles (QP), and Purcell. The losses are calculated for qubit mode temperature of $T_Q = 45$ mK and resonator mode temperature of $T_R=50$ mK. We refer the reader to appendix \ref{app:decay} for the exact dissipation model and its temperature dependence.}
\end{figure}

When the absolute value of the dispersive shift $\chi_{QC}$ is high, the storage mode coherence time will still be limited by the ancilla. For this system to be suitable as a bosonic qubit, further understanding of the fluxonium losses is required. We measure the lifetime of the fluxonium qubit $T_1^Q$ as a function of the flux bias and plot the $T_1^Q$ limit from the dominant decay channels\cite{pop_coherent_2014, Clark_2020, Clerk_review_2010}: dielectric loss, inductive loss, Purcell limit\cite{Zhang_Purcell_fluxonium_2021}, and quasiparticle loss\cite{catelani_quasiparticle_2011}
(see Fig. \ref{fig:t1}). From the fitted bounds for the quality factors of each dissipation channel, we deduce that the lifetime of the fluxonium at the zero field spot is limited by dielectric losses. Besides the range where the qubit is close to resonant with the readout resonator, it is evident that the qubit is also Purcell limited\cite{purcell_spontaneous_1946, Hanai_purcell_2021} at $\Phi_\text{ext}/\Phi_0 = 0.5$. These limits are further reduced at this flux bias when including the high thermal population of both the qubit and the resonator. The system would greatly benefit from a Purcell filter and improved fabrication techniques. A possible improvement in the fabrication would be to treat the sample with hydrofluoric (\text{HF}) acid after fabrication, which has been shown to reduce dielectric losses due to two-level system defects in resonators on silicon\cite{chayanun_characterization_2024}. Another possibility is to adapt the fabrication process so it is compatible with more aggressive surface cleaning by swapping the \text{Al} with either tantalum\cite{place_new_2021,wang_towards_2022} or niobium capped by another metal\cite{bal_systematic_2024}.

\section{Conclusion and Outlook}
To summarize, we have demonstrated that a fluxonium is a viable ancilla to a high-Q coaxial cavity. We explore for the first time that this platform realizes tunable interaction with an invertible sign without using supplementary circuit elements like tunable couplers or applying parametric drives. This appends another element to the toolbox to control higher coherence microwave cavities, which could enable novel control strategies for building bosonic codes. The fact that the ancilla can be decoupled from the bosonic mode alleviates the stringent requirement for a high-coherence ancilla without requiring a trade-off with the bosonic qubit control capabilities. 

Further improvements of the systems involve including a Purcell filter which would greatly increase the lifetime of the fluxonium. In particular, the intrinsic Purcell protection\cite{Nakamura_purcell_2022} is compatible with this setup. Adding fast flux control to the flux hose will allow fast, in-situ switching between the different interaction regimes. Such a system would be ideally suited for pursuing new quantum gates or creating and evacuating quantum states from the long-lived memory mode.

\begin{acknowledgments}

D.G.A. and I.Y. would like to thank S. Oleschko for the guidance during the flux hose fabrication. We thank P. Groszkowski, J. Koch, and S. P. Chitta for their valuable input on the quasiparticle losses. 

This research was funded in whole or in part by the Austrian Science Fund (FWF), Grant DOI 10.55776/W1259, and the 10.55776/I4395 QuantERA grant QuCOS. T.H.-D. is funded by the 10.55776/COE1 QuantA grant. For open access purposes, the author has applied a CC BY public copyright license to any author-accepted manuscript version arising from this submission.

\end{acknowledgments}

\section*{Data Availability Statement}

The data and code used to support the findings of this study are available on Zenodo.

\appendix
\begin{figure*}
\includegraphics{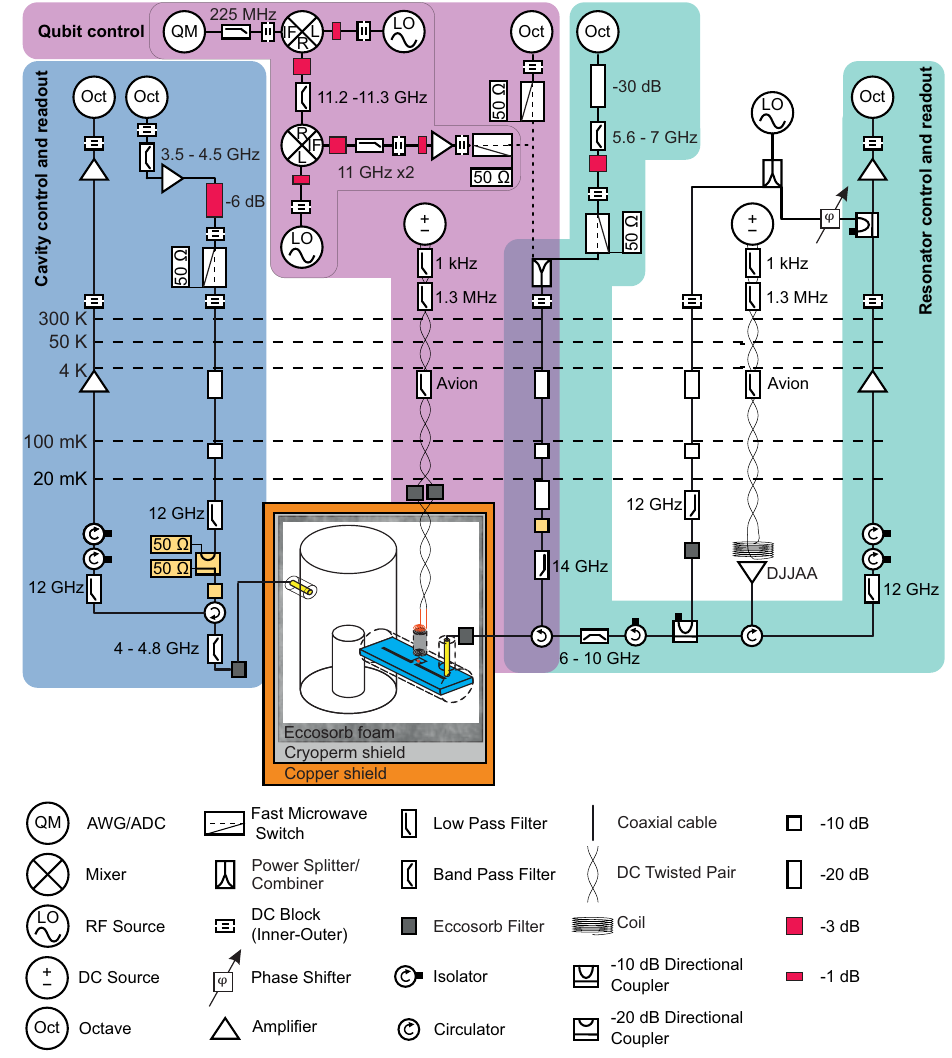}
\caption{\label{fig:wiring} Wiring diagram of the experiment. The lined part of the qubit control corresponds to the low-frequency microwave setup. The dashed lines refer to the different plates of the cryostat. Depending on the measurement, either the low-, or high-frequency branch, or both branches are combined with the readout tone. The elements highlighted in yellow are specially thermalized to the base plate. The components related only to the parametric amplifier are not highlighted.}
\end{figure*}
\section{Experimental Setup}

A schematic of the cryogenic and microwave setup is shown in Fig. \ref{fig:wiring}. The measurements were conducted in an Oxford Triton DU7-200 cryofree dilution refrigerator system.

The radio frequency (RF) input coaxial lines were attenuated by $20$ dB at the $4$K plate and $ 10$ dB at the still plate. On the base plate, the input signal for the qubit line was attenuated further by $30$ dB (a $20$ dB attenuator and a  Quantum Microwave $10$ dB thermalized attenuator) and then filtered with a Mini-Circuits DC-14 GHz low pass filter. The experiment was in a reflection configuration and a Low Noise Factory single junction $4-12$ GHz circulator was used. The signal was finally routed through a home-built infrared filter. The input signal for the cavity has the same configuration up to the $100$ mK stage. A low pass K\&L filter filters the cavity signal which is then attenuated by a  Quantum Microwave $20$ dB thermalized directional coupler and $10$ dB attenuator from the same brand. A narrow-band Mini-Circuits $4 - 4.8$ GHz band-pass filter and an infrared filter are placed just outside the shield. A  Quinstar $4-8$ GHz circulator allows for direct reflection measurements of the cavity.

The sample was thermalized to the base plate via copper clamps. The experiment was shielded by a µ-metal shield and enclosed by a copper can, covered with eccosorb paint on the inside. In the inner can, the sample was surrounded by eccosorb foam. 

The output signal from the resonator was filtered by a $6-10$ GHz Microtronics band-pass filter and then routed through an isolator, a Krytar $10$ dB directional coupler, and a Low Noise Factory circulator before reaching a parametric DJJAA (a Dimer Josephson Junction Array Amplifier\cite{winkel_djjaa_2020}). The $10$ dB directional coupler is used to couple the amplifier pump line. After the parametric amplifier, the output signal goes through two Quinstar isolators before being amplified at the $4$K stage by a high electron mobility amplifier and again at room temperature. To avoid amplifier saturation, an additional $10$ dB directional coupler before the room-temperature amplifier is added to cancel the DJJAA pump tone from the fridge. 

A Quantum Machine Operator X+ (OPX+) was used together with a Quantum Machine Octave to generate and digitize the microwave signals. The OPX+ has arbitrary waveform generators that produce the pulses used in the experiment, which are then mixed up to the relevant frequencies and amplified in the Octave. For readout, the Octave down-mixes the resonator and cavity pulses and is digitized by the OPX+. In the low-frequency range $f\in[1,6.8]$ GHz
the fluxonium pulses were up-mixed via a double-super-heterodyne \cite{horowitz_textbook_CUP_1989} scheme that employs two local oscillators and two single side-band mixers to produce a wide range of frequencies without the need for calibration. For higher qubit frequencies, the pulses are directly mixed by Octave.  The pulse generation setup includes other amplifiers, filters fast microwave switches, and DC blocks to produce a high amplitude of the wanted microwave signal, reduce leakage of unwanted signals, and prevent ground loops. The flux biasing is done by leading in DC signals via twisted pairs through commercial low-pass filters. Two home-built high-attenuation filters further refine the qubit bias.

\section{Fabrication}\label{app:fab}
The fabrication recipe was adapted from \cite{gusenkova_quantum_2021} for the University of Innsbruck Quantum Nano-Zentrum Tirol (QNZT) cleanroom. The fluxonium and resonator were patterned by electron-beam writer (Raith eLINE Plus 30kV) on a bi-layer resist stack ($0.96$ µm MMA (8.5) EL13 and $0.24$ µm of 950 PMMA A4). The substrate wafer used is a 2-inch sapphire wafer that was piranha-cleaned before processing. To prevent the charging of the substrate, an $8$ nm thin aluminum layer is evaporated onto the surface of the resist stack. After lithography, the aluminum layer was etched in a TMAH-based developer, and the resist was developed in a 3:1 solution of isopropyl alcohol (IPA) and deionized (DI) water. Next, 20 nm \text{Al}, 30 nm \text{Al}, and 40 nm \text{grAl}  layers were evaporated onto the sample using a Plassys MEB550S electron-beam evaporator. The \text{Al} layers were evaporated under an angle with a controlled oxidation step ($29.6$ mbar for $5$ min) in between the deposition to create the Dolan-bridge junction. The \text{grAl} is deposited by evaporating \text{Al} under an oxygen flow. Before venting the chamber, the wafer was exposed to a pure oxygen environment for $10$ minutes at $10$ mbar to avoid impurities in the aluminum oxide capping layer on top of the sample. The resist layer was lifted off in N-Methyl-2-pyrrolidone (NMP) at 80$^{\circ}$C and cleaned in acetone and IPA. Finally, the samples were laser-diced with a Coherent EXACTCUT 430 laser dicer.

Immediately after machining, the post cavity has a post length of $14.8$ mm, an inner radius of $2.2$ mm, and an outer radius of $5.8$ mm. The tunnel for the sapphire chip has a diameter of $3.8$ mm and is aligned with the post height to maximize the coupling between the high Q cavity mode and the qubit. The high-purity aluminum cavity was machined at the Institute for Quantum Optics and Quantum Information Innsbruck mechanical workshop via an electro-discharge machining method. After manufacturing, the cavity was etched in two rounds with Transene aluminum etchant type A that removes approximately $150$ µm of material. Finally, the cavity is thoroughly washed with DI water and dried. 
\section{Simulations}\label{app:sim}
The high nonlinearity of the fluxonium in combination with distributed elements coupled to it makes the system hard to simulate. Finite element simulators, such as Ansys HFSS, in conjunction with either BBQ or the Qiskit Metal extension: pyEPR\cite{Minev_epr_2021}, have until now failed to simulate highly anharmonic qubits. As we have shown, scQubits provide the possible solution but a few subtleties need to be taken care of. 

\begin{figure}[h!]
 \centering
\includegraphics{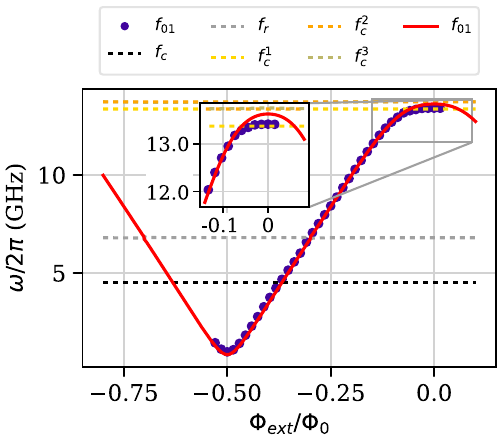}
\caption{\label{fig:extra_modes} Fluxonium spectroscopy with simulation curves including higher cavity modes. In the model, apart from the fundamental modes of the cavity $f_c$ and the resonator $f_r$, three higher cavity modes are included: $f_c^1 = 13.375$ GHz, $f_c^2=13.735$ GHz, and $f_c^3 = 13.734$ GHz. They are capacitively coupled to the resonator with rates $g_{\text{RC}^{1,2,3}}/2\pi=16$ MHz. Moreover, the highest frequency mode at $f_c^3$ is directly coupled to the qubit. This mode does not affect the frequencies significantly even to a direct coupling strength of $g_{\text{QC}^3}/2\pi=16$ MHz. Inset: zoom-in of the region around $\Phi_{ext}/\Phi_0=0$.}
\end{figure}

Firstly, all interaction terms are required as rotating wave approximation is not applicable in order to get the correct interaction strengths. Secondly, the higher cavity modes cannot be neglected and will influence the bare frequencies of the harmonic oscillators, needed for the fit. They will also push the qubit resonance to lower frequencies than anticipated. In Fig. \ref{fig:extra_modes} we show the influence of higher cavity modes on the qubit fundamental frequency. Three higher modes of the cavity are included with direct coupling to the resonator, the highest frequency mode also is coupled directly to the qubit. The coupling strengths were based on expectation values from the scaling with $\sqrt{\omega}$ and then further increased to amplify their effect. As the frequencies of those modes were never measured, but extracted from simulations, the actual cavity modes could coincide with the qubit levels, which would cause a further shift in the transition $f_{01}$. Such a coincidence could explain why around $\Phi_\text{ext}/\Phi_0=0$ the measured qubit frequencies fit better to the cavity mode at $f_c^1$.

The last remark concerns the fitting routine. Initially, the nonlinear optimization and curve fitting of the qubit resonance as a function of flux were executed with a Python solver lmfit. However, the energies from the initial fit were adapted from $E_J/h=10.34\pm0.04$ GHz, $E_C/h=3.36\pm 0.04$ GHz and $E_L=1.038\pm 0.003$ GHz to the values from the main text to accommodate for the avoided crossing in Fig. \ref{fig:disp_shifts}b, observed around the high sweet spot. Upon further investigation,  the cause of this crossing is the two-photon qubit transition $f_{13}$ crossing the cavity resonance. The exact location of the higher energy levels plays an important role in how $\chi$ behaves. Even small changes in the qubit energies $E_J,\, E_C,\, E_L$ result in new features, such as the aforementioned avoided crossing. For the most accurate predictions of the dispersive shift, the fit must include the higher fluxonium transitions. 

\section{Qubit decay channels}\label{app:decay}
In this section, we discuss in detail the limitations on the qubit lifetime via the various qubit decay channels\cite{pop_coherent_2014, Clark_2020, Hanai_purcell_2021}. The decay rate of the qubit is given by Fermi's Golden Rule\cite{Clerk_review_2010} of the type
\begin{equation}
    \Gamma_{01}=\frac{1}{\hbar^2}|\langle 0|\hat{\mathcal{O}}|1\rangle|^2[S(\omega_{01})+S(-\omega_{01})].
\end{equation}
The operator $ \hat{\mathcal{O}} $ is the qubit operator which couples to the lossy bath and $S(\omega_{01})$ is the noise spectral density at the qubit frequency. The first term corresponds to emission into the bath, and the second for absorption. 

We consider the most common loss mechanisms. For dielectric loss, the flux operator $\hat{\mathcal{O}}=\phi_0\hat{\varphi}$ couples to the bath voltage with density 
\begin{equation}
    S_{VV}(\omega_{01})+S_{VV}(-\omega_{01})=\frac{2\hbar\omega_{01}^2C_J}{Q_{\text{diel}}(\omega)}\text{coth}\left(\frac{\hbar|\omega|}{2k_BT}\right)
\end{equation}
where $C_J=e^2/2E_C$ is the junction capacitance, $T$ is the qubit mode temperature, and $k_B$ - Boltzmann's constant. The quality factor of the dielectric losses is weakly frequency-dependant
\begin{equation}
    Q_{\text{diel}}(\omega) = Q_{\text{diel}}\left(\frac{2\pi\, 6\,\text{GHz}}{|\omega|}\right)^{0.7}.
\end{equation}
Here, the $Q_{\text{diel}}$ is the nominal quality factor quoted for $6$ GHz.

In the case of inductive losses, the flux operator $\hat{\mathcal{O}}=\phi_0\hat{\varphi}$ couples to the bath current with density 
\begin{equation}
     S_{II}(\omega_{01})+S_{II}(-\omega_{01})=\frac{\hbar}{LQ_{\text{ind}}}\,\text{coth}\left(\frac{\hbar|\omega|}{2k_BT}\right)
\end{equation}
where $L = \phi_0^2/E_L$ is the fluxonium superinductance and $ Q_{\text{ind}}$ is the inductive quality factor.

The quasiparticle loss would result from the qubit operator $\hat{\mathcal{O}}=2\phi_0\,\text{sin}\frac{\hat{\varphi}}{2}$ coupling to a noise variable with density
\begin{equation}
    S_{\text{qp}}(\omega)+S_{\text{qp}}(-\omega)=2\hbar\omega\,\Re \{Y_{\text{qp}}(\omega)\}\text{coth}\left(\frac{\hbar|\omega|}{2k_BT}\right)
\end{equation}
with the dissipative part of the Josephson junction admittance having the form
\begin{eqnarray}
    \Re \{Y_{\text{qp}}(\omega)\}=&&\sqrt{\frac{2}{\pi}}\frac{8E_J}{R_k\Delta_{\text{Al}}}\left(\frac{2\Delta_\text{Al}}{\hbar\omega}\right)^{1.5}x_{\text{qp}}\sqrt{\frac{\hbar\omega}{2k_BT}}\nonumber\\
    &&\times K_0\left(\frac{\hbar|\omega|}{2k_BT}\right)\text{sinh}\left(\frac{\hbar|\omega|}{2k_BT}\right).
\end{eqnarray}
Here, $R_k = h/e^2$ is the resistance quantum, and $\Delta_\text{Al}=1.76T_c^\text{Al}k_B$  the $\text{Al}$ superconducting gap with critical temperature $T_c^\text{Al}$. The fitted value here is the quasiparticle density $x_{\text{qp}}$ relative to the number of Cooper pairs. For Fig. \ref{fig:t1}, the qubit operator has been transformed to $\hat{\mathcal{O}}=\phi_0\,\text{sin}(\frac{\hat{\varphi}}{2}+\frac{\hat{I}}{2})$, where $\hat{I}$ is the identity operator.
\begin{figure}[h!]
 \centering
\includegraphics{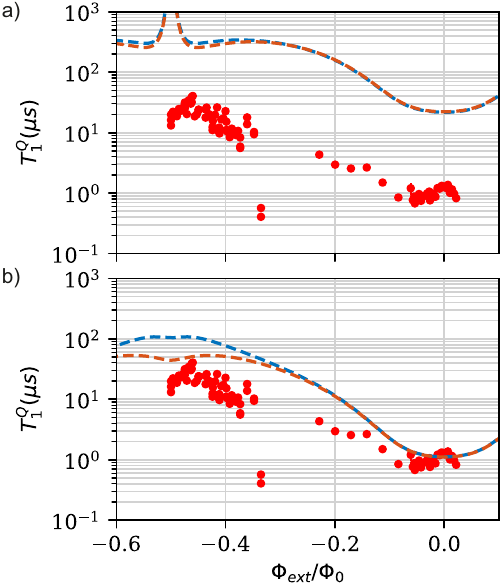}
\caption{\label{fig:diel} Temperature dependence of the quasiparticle and dielectric losses for $45$ (blue) and $112$ mK (red) qubit mode temperature. a) Quasiparticle losses for quasiparticle density $x_{\text{qp}}= 1.25\cdot 10^{-6}$. b) Dielectric losses for quality factor $Q_{\text{diel}}=1.5\cdot 10^5$. The biggest influence the temperature has is around $\Phi_\text{ext}/\Phi_0=-0.5$. Only the respective loss channel for each plot is considered and all other loss contributions are neglected.}
\end{figure}

The Purcell loss was calculated according to the supplementary material in \cite{Zhang_Purcell_fluxonium_2021}. The absorption rate from the bath of oscillators through the readout resonator is
\begin{equation}
\Gamma_{|0\rangle_Q\rightarrow|1\rangle_Q}^{\text{Purcell}\uparrow} = P_{\text{res}}(0) \kappa n_{\text{th}}(\omega_{01})|\langle01|\hat{a}^\dagger|00\rangle|^2
\end{equation}
while the emission rate into the bath is
\begin{equation}
\Gamma_{|1\rangle_Q\rightarrow|0\rangle_Q}^{\text{Purcell}\downarrow} = P_{\text{res}}(0) \kappa [n_{\text{th}}(-\omega_{01})+1]|\langle00|\hat{a}|01\rangle|^2.
\end{equation}
Here, $P(n) = [1-e^{-\hbar\omega_R/k_BT_R}]e^{-n\hbar\omega_R/k_BT_R}$ is the probability to have $n$ photons in a resonator with mode temperature $T_R$, $\kappa$ is the total loss rate of the resonator, and $n_{\text{th}}(\omega) = \frac{1}{e^{\hbar\omega/k_BT}-1}$. The states $|00\rangle$ correspond to the dressed state of the coupled qubit-resonator system with the biggest overlap to $|0\rangle_R|0\rangle_Q$ in the bare basis.

\begin{figure}[h!]
 \centering
\includegraphics{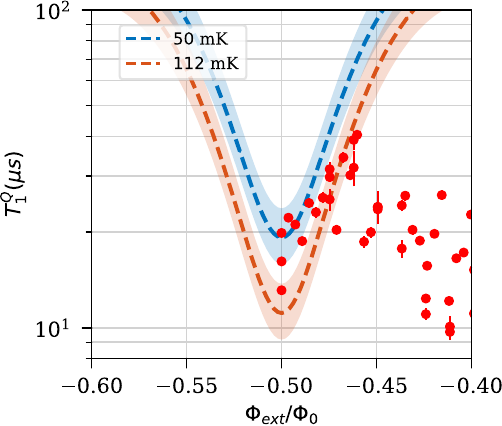}
\caption{\label{fig:purcell} Adjusted Purcell limit. The temperature of the resonator mode is adjusted from $112$ to $50$ mK to show the temperature change during the measurements. The shaded regions correspond to a change in the coupling rate $g_\text{QR}$ of $\pm10\%$ for each resonator temperature. All other loss mechanisms are neglected.}
\end{figure}

For the fit in Fig. \ref{fig:t1}, a qubit mode temperature of $T_Q = 45$ mK, a nominal resonator mode temperature $T_R=50$ mK, and the total loss rate at the low sweet spot $\kappa_R/2\pi=922$ kHz were used. However, the qubit mode temperature obtained before the sample was fully thermalized to the base plate of the cryostat is $T_Q = 112\pm2$ mK. The minimum measured qubit temperature is $45$ mK five months after the cooldown. In Fig. \ref{fig:diel}, the effect of qubit mode temperature on the quasiparticle and dielectric losses is shown. Depending on the qubit mode temperature, the quasiparticle density (Fig. \ref{fig:diel}a) and the dielectric quality factors (Fig. \ref{fig:diel}b) need adjustment to fit the same qubit lifetimes. From a qubit spectroscopy measurement, we set the upper limit on the resonator mode temperature to $T_R = 112 \pm 29$ mK. For this temperature the Purcell limit set on the qubit lifetime lies below the measured data. As the biggest contributors to the Purcell decay around $\Phi_\text{ext}/\Phi_0 = 0.5$ are the coupling strength $g_\text{QR}$ and the resonator temperature, we vary both, as shown in Fig. \ref{fig:purcell}. The lower temperature of 50 mK is chosen to be slightly above the minimum measured qubit mode temperature, while the variation of the coupling strength corresponds to 10\% of its nominal value (refer to Table \ref{tab:system_parameters}). Even with reduced interaction strength, the Purcell limit at $T_R=112$ mK lies below the measured fluxonium lifetimes, which dictates our choice to use $T_R = 50$ mK instead. For the lower resonator mode temperature and reduced coupling strength, the qubit lifetime is fully Purcell-limited around the low sweet spot.

\begin{figure}[h!]
 \centering
\includegraphics{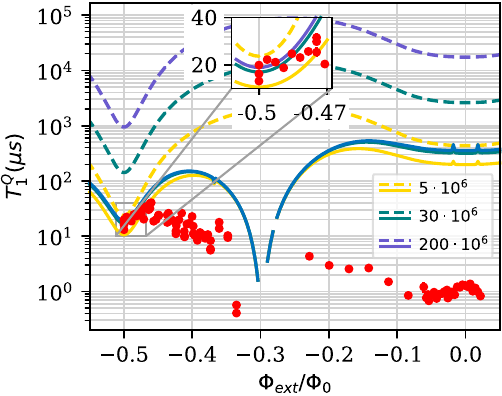}
\caption{\label{fig:ind} Limit for Purcell decay and inductive losses with varying inductive quality factor. The solid line represents the sum of inductive and Purcell loss, while the dashed line is only the limit set by inductive loss. The inset focuses on the illustrated losses around the low sweet spot.}
\end{figure}

To justify the limits we have set on the inductive losses, we plot the bound by both Purcell decay and inductive loss with varying quality factors (Fig. \ref{fig:ind}). The biggest impact of these dissipation channels is around the low sweet spot. When dielectric loss is included, the total bound is lowered further, imposing a high inductive quality factor to fit the qubit lifetimes around $\Phi_\text{ext}/\Phi_0=-0.5$. 

\section{Measurement methods}\label{app:methods}
The dispersive shift between the resonator and the fluxonium is much higher than both the linewidth of the resonator and the qubit, allowing for easy extraction of the quantity as a function of flux bias. For Fig. \ref{fig:disp_shifts}a, several approaches were used. Firstly, because of the high mode temperature of the qubit, the excited state population is non-negligible around $\Phi_\text{ext}/\Phi_0=0.5$, causing the resonator response to split by an amount equal to the dispersive shift. From the depth ratio of the two resonances and the measured qubit frequency, the qubit mode temperature was calculated. At higher qubit frequencies, the qubit cannot be thermally excited by the bath and the resonator has a single resonance. At these flux bias points a two-tone measurement was needed, in which an extra tone with varying frequency around the qubit frequency was applied, while the resonator spectroscopy was monitored. Once the frequency was correct, the resonance splitting was noted. At the high sweet spot $\Phi_\text{ext}/\Phi_0=0$, $\chi_{QR}$ was extracted from qubit spectroscopy after pulsing the resonator.

Despite the dispersive shift $\chi_{QC}$ being smaller than the fluxonium decoherence rate, using large displacement pulses on the cavity enhances the system response, making it possible to extract the Hamiltonian parameters. This technique was implemented in \cite{Phillip_2020,eickbusch_ECD_2022}. When the cavity is displaced to $\alpha_0$, due to its coupling to the qubit, the coherent state will start rotating in phase space around the origin. The rotation frequency will depend on the qubit state: $\Delta_g = \Delta - 2K_C\alpha_0^2$ and $\Delta_e=\Delta-\chi_{QC}-(2K_C+2\chi_{nl})\alpha_0^2$, where $\Delta$ is the detuning between the cavity drive and the mode frequency, $K_C$ is the Kerr shift of the cavity and $\chi_{nl}$ is a higher-order dispersive shift.

\begin{figure}[h!]
 \centering
\includegraphics{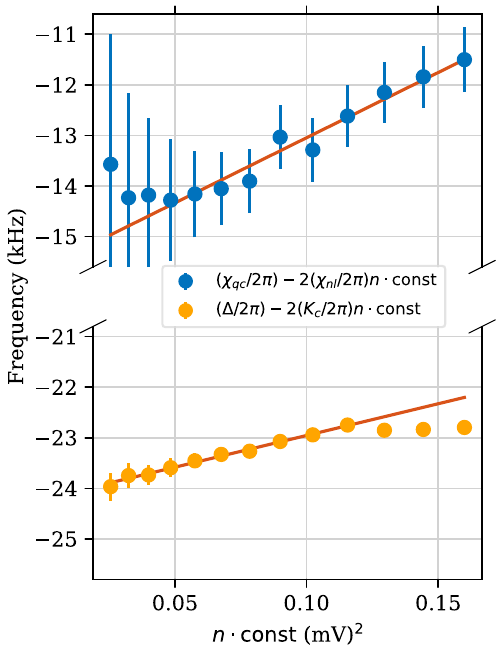}
\caption{\label{fig:chis} Cavity frequency dispersion at $\Phi_\text{ext}=0.5\Phi_0$ as a function of the pulse voltage squared. The rotation frequency $\Delta_g/2\pi$ is plotted in yellow, while the difference $(\Delta_e-\Delta_g)/2\pi$ is plotted in blue. The data sets are fitted by linear regression (red). The voltage squared is proportional to the number of photons in the cavity up to a constant. From the fit, we extract $\Delta/2\pi=-24.21\pm 0.03$ kHz and $\chi_{QC}/2\pi=-15.6\pm 2$ kHz, the latter of which we use to find the unknown constant and thus calibrate the photon number $n$.}
\end{figure}

By applying a selective $1.2\,\mu s$ Gaussian pulse, we prepare the qubit in the ground/excited state with the help of the parametric amplifier. The excited state population in steady state is around $36.5\%$, after preparation in ground state it is reduced to $27.7\%$. The preparation of the qubit in an excited state with active reset gives us $73.5\%$ excited state population, compared to $51.5\%$ without feedback from the parametric amplifier. The preparation fidelity is limited by the qubit lifetime and the external coupling rate of the readout resonator. Following the state preparation, we displace the cavity with a $40$ ns square pulse. After a fixed time of $1\,\mu$s, we try to displace the cavity back to the origin by applying a pulse with varying phases. We repeat this out-and-back procedure 5 times to enhance the measurement sensitivity. At the end of the protocol, we flip the qubit with a narrow-bandwidth $\pi$-pulse and then read it out with a $2\,\mu s$ square pulse. The $\pi$-pulse is only successful if the phase for the displacement back to the cavity ground state is correct. We repeat the procedure for different displacements $\alpha_0$. The correct phase of the cavity pulse is then converted to frequency. We fit the data for $\Delta_e-\Delta_g$ to a linear regression as shown in Fig. \ref{fig:chis}, where the intercept gives us the dispersive shift at the low sweet spot $\chi_{QC}/2\pi = -15.6\pm 2$ kHz. Similarly, the detuning can be determined from fitting $\Delta_g$, from where we extract $\Delta/2\pi = -24.21\pm 0.03$ kHz.

To extract the slope parameters ($K_C$ and $\chi_{nl}$), the cavity drive strength is calibrated. We prepare the qubit in a superposition state by applying a $\pi/2$-pulse. We then apply a sequence similar to the echoed conditional displacement\cite{eickbusch_ECD_2022}, without the $\pi$-pulse in between, i.e. $\hat{D}(\alpha)\hat{T}(t_w)\hat{D}(-r\alpha)\hat{D}(-r\alpha)\hat{T}(t_w)\hat{D}(\alpha)$, where $\hat{D}(\alpha)=e^{\alpha\hat{a}-\alpha^*\hat{a}^\dagger}$ is the displacement by $\alpha$, $\hat{T}(t_w)$ allows the system to evolve for time $t_w=200$ ns, and $r$ is correction ratio to make pulses more accurate. During the sequence, the cavity and qubit are entangled. After disentangling the two subsystems, the trajectories for the ground and excited qubit states will have acquired different phases. We measure this phase by either applying $\hat{X}_{\pi/2}$ or $\hat{Y}_{\pi/2}$ and reading out the qubit population $\langle\sigma_x\rangle$ and $\langle\sigma_y\rangle$. We repeat the sequence for different cavity pulse amplitudes. To fit the curves, the semi-classical trajectories are numerically simulated, allowing us to extract the enclosed area for each drive power and convert it to the accumulated phase difference, which we later fit to the measured $\langle\sigma_x\rangle$ and $\langle\sigma_y\rangle$. This gives us a conversion factor $\sqrt{\text{const}}$ from Fig. \ref{fig:chis} between pulse voltage and the displacement we can achieve. For more details on the exact sequence and the semi-classical trajectories, we refer the reader to the supplementary material of \cite{eickbusch_ECD_2022}.

As this method is time-consuming, only the dispersive shift at $\Phi_\text{ext}/\Phi_0=0.5$ was measured this way. As the cavity frequency does not change much as a function of flux threading the fluxonium loop, we use the same photon number calibration from the measured dispersive shift for the rest of the points in Fig. \ref{fig:disp_shifts}b. These points were extracted by monitoring the qubit frequency after the cavity had been displaced to a different coherent state. We fit the qubit resonance shift observed as a function of the photon number and extract the dispersive shift $\chi_{QC}/2\pi$. These subsequent measurements allowed for fast characterization of the dispersive shift across the entire flux map.

With the gained knowledge of the system, the lifetime of the storage mode was measured. We displace the cavity and let it decay while recording the characteristic function as we wait (details can be found in the supplementary material of \cite{Phillip_2020}). The coherent state will decay from the cavity as $\alpha(t)=\alpha_0e^{-i\Delta t-\frac{t}{2T_1^C}}$. The characteristic function of such a coherent state in the cavity is described as $\mathcal{C}(\beta,t) = e^{\frac{|\beta|^2}{4}}e^{-i\Im\{\alpha(t)\beta^*\}}$, where $\beta$ determines the point of phase space where the function is measured. From $\mathcal{C}(\beta,t)$ for $\beta\in\mathbb{R}$, we extract $\alpha(t)$ at each recorded time and fit it to extract the cavity lifetime. This procedure was only feasible at $\Phi_\text{ext}=-0.5$ as the fluxonium coherence time was insufficient at other flux bias points.

\begin{table*}
\caption{\label{tab:system_parameters} System Parameters. The cavity frequency $\omega_c/2\pi$ has additional uncertainty of 400 kHz coming from the 15\% variation in $g_\text{RC}$. Due to the dephasing at $\Phi_\text{ext}/\Phi_0=0.5$ caused by the excited qubit population, the cavity coupling rate and linewidth were not measured at this flux bias.}

\begin{tabular}{||c|c|c||}
    \hline
    Parameter & Symbol & Value \\ 
     \hline\hline

     Josephson Energy& $E_J/h$ & $10.8$ GHz \\
     Inductive Energy & $E_L/h$ & $1.014$ GHz \\
     Capacitive Energy & $E_C/h$ & $3.5$ GHz \\
     
     Bare High Q Cavity Frequency& $\omega_{\mathrm{c}}/2\pi$ & $4.535854$ GHz \\
     
     Bare Readout Resonator Frequency & $\omega_{\mathrm{r}}/2\pi$ & $6.8176$ GHz \\
     
     Qubit-Resonator Coupling & $g_\text{QR}/2\pi$ & $25.2$ MHz $\pm 10\%$\\
     
     Resonator-High Q Cavity Coupling & $g_\text{RC}/2\pi$ & $8$ MHz$\pm 15\%$\\
     \hline
     High Q Cavity Lifetime at $\Phi_\text{ext}/\Phi_0=0.5$  & $T_1^C$ & $210\pm 40$ µs \\

     Qubit Residual Mode Temperature at $\Phi_\text{ext}/\Phi_0=0.5$ & $T_{\mathrm{Q}}$ & $112 \pm 2$ mK \\ 
      Qubit Excited State Population at $\Phi_\text{ext}/\Phi_0=0.5$ & $P_e$ & $39.1 \pm 0.7$ $\%$ \\ 
     Resonator Residual Mode Temperature at $\Phi_\text{ext}/\Phi_0=0.5$ & $T_{\mathrm{R}}$ & $110\pm30$ mK \\
     Readout Resonator Linewidth at $\Phi_\text{ext}/\Phi_0=0.5$& $\kappa_R/2\pi$ & $922\pm 3$ kHz\\
     Readout Resonator External Coupling Rate at $\Phi_\text{ext}/\Phi_0=0.5$& $\kappa^\text{ext}_R/2\pi$ & $549\pm 2$ kHz\\
       Readout Resonator Linewidth at $\Phi_\text{ext}/\Phi_0=0$& $\kappa_R/2\pi$ & $956\pm 4$ kHz\\
     Readout Resonator External Coupling Rate at $\Phi_\text{ext}/\Phi_0=0$& $\kappa^\text{ext}_R/2\pi$ & $801\pm2$ kHz\\
     Cavity Linewidth at $\Phi_\text{ext}/\Phi_0=0$& $\kappa_C/2\pi$ & $550\pm20$ Hz \\
     Cavity External Coupling Rate at $\Phi_\text{ext}/\Phi_0=0$& $\kappa^\text{ext}_C/2\pi$ & $233\pm7$ Hz \\

     \hline
     Dielectric quality factor at $6$ GHz & $Q_{\text{diel}}$ & $\gtrsim0.15 \cdot 10^6$\\
     Inductive quality factor & $Q_{\text{ind}}$ & $\gtrsim30 \cdot 10^6$\\
    Non-equilibrium quasi-particle density & $x_{\text{qp}}$ & $\lesssim 1.25\cdot 10^{-6}$\\
    \hline
\end{tabular}
\end{table*}

\bibliography{aipsamp}

\end{document}